\documentclass[aps,prl,showpacs,twocolumn,superscriptaddress,groupedaddress,nobibnotes]{revtex4}
\usepackage{graphicx}
\begin{document}

\widetext
\hspace{5.2in} \mbox{Fermilab-Pub-14-036}

\title{Study of Top-Quark Production and Decays involving a
       Tau Lepton at CDF and Limits on a Charged-Higgs Boson
                             Contribution

                                                       }

\affiliation{Institute of Physics, Academia Sinica, Taipei, Taiwan 11529, Republic of China}
\affiliation{Argonne National Laboratory, Argonne, Illinois 60439, USA}
\affiliation{University of Athens, 157 71 Athens, Greece}
\affiliation{Institut de Fisica d'Altes Energies, ICREA, Universitat Autonoma de Barcelona, E-08193, Bellaterra (Barcelona), Spain}
\affiliation{Baylor University, Waco, Texas 76798, USA}
\affiliation{Istituto Nazionale di Fisica Nucleare Bologna, \ensuremath{^{ii}}University of Bologna, I-40127 Bologna, Italy}
\affiliation{University of California, Davis, Davis, California 95616, USA}
\affiliation{University of California, Los Angeles, Los Angeles, California 90024, USA}
\affiliation{Instituto de Fisica de Cantabria, CSIC-University of Cantabria, 39005 Santander, Spain}
\affiliation{Carnegie Mellon University, Pittsburgh, Pennsylvania 15213, USA}
\affiliation{Enrico Fermi Institute, University of Chicago, Chicago, Illinois 60637, USA}
\affiliation{Comenius University, 842 48 Bratislava, Slovakia; Institute of Experimental Physics, 040 01 Kosice, Slovakia}
\affiliation{Joint Institute for Nuclear Research, RU-141980 Dubna, Russia}
\affiliation{Duke University, Durham, North Carolina 27708, USA}
\affiliation{Fermi National Accelerator Laboratory, Batavia, Illinois 60510, USA}
\affiliation{University of Florida, Gainesville, Florida 32611, USA}
\affiliation{Laboratori Nazionali di Frascati, Istituto Nazionale di Fisica Nucleare, I-00044 Frascati, Italy}
\affiliation{University of Geneva, CH-1211 Geneva 4, Switzerland}
\affiliation{Glasgow University, Glasgow G12 8QQ, United Kingdom}
\affiliation{Harvard University, Cambridge, Massachusetts 02138, USA}
\affiliation{Division of High Energy Physics, Department of Physics, University of Helsinki, FIN-00014, Helsinki, Finland; Helsinki Institute of Physics, FIN-00014, Helsinki, Finland}
\affiliation{University of Illinois, Urbana, Illinois 61801, USA}
\affiliation{The Johns Hopkins University, Baltimore, Maryland 21218, USA}
\affiliation{Institut f\"{u}r Experimentelle Kernphysik, Karlsruhe Institute of Technology, D-76131 Karlsruhe, Germany}
\affiliation{Center for High Energy Physics: Kyungpook National University, Daegu 702-701, Korea; Seoul National University, Seoul 151-742, Korea; Sungkyunkwan University, Suwon 440-746, Korea; Korea Institute of Science and Technology Information, Daejeon 305-806, Korea; Chonnam National University, Gwangju 500-757, Korea; Chonbuk National University, Jeonju 561-756, Korea; Ewha Womans University, Seoul, 120-750, Korea}
\affiliation{Ernest Orlando Lawrence Berkeley National Laboratory, Berkeley, California 94720, USA}
\affiliation{University of Liverpool, Liverpool L69 7ZE, United Kingdom}
\affiliation{University College London, London WC1E 6BT, United Kingdom}
\affiliation{Centro de Investigaciones Energeticas Medioambientales y Tecnologicas, E-28040 Madrid, Spain}
\affiliation{Massachusetts Institute of Technology, Cambridge, Massachusetts 02139, USA}
\affiliation{University of Michigan, Ann Arbor, Michigan 48109, USA}
\affiliation{Michigan State University, East Lansing, Michigan 48824, USA}
\affiliation{Institution for Theoretical and Experimental Physics, ITEP, Moscow 117259, Russia}
\affiliation{University of New Mexico, Albuquerque, New Mexico 87131, USA}
\affiliation{The Ohio State University, Columbus, Ohio 43210, USA}
\affiliation{Okayama University, Okayama 700-8530, Japan}
\affiliation{Osaka City University, Osaka 558-8585, Japan}
\affiliation{University of Oxford, Oxford OX1 3RH, United Kingdom}
\affiliation{Istituto Nazionale di Fisica Nucleare, Sezione di Padova, \ensuremath{^{jj}}University of Padova, I-35131 Padova, Italy}
\affiliation{University of Pennsylvania, Philadelphia, Pennsylvania 19104, USA}
\affiliation{Istituto Nazionale di Fisica Nucleare Pisa, \ensuremath{^{kk}}University of Pisa, \ensuremath{^{ll}}University of Siena, \ensuremath{^{mm}}Scuola Normale Superiore, I-56127 Pisa, Italy, \ensuremath{^{nn}}INFN Pavia, I-27100 Pavia, Italy, \ensuremath{^{oo}}University of Pavia, I-27100 Pavia, Italy}
\affiliation{University of Pittsburgh, Pittsburgh, Pennsylvania 15260, USA}
\affiliation{Purdue University, West Lafayette, Indiana 47907, USA}
\affiliation{University of Rochester, Rochester, New York 14627, USA}
\affiliation{The Rockefeller University, New York, New York 10065, USA}
\affiliation{Istituto Nazionale di Fisica Nucleare, Sezione di Roma 1, \ensuremath{^{pp}}Sapienza Universit\`{a} di Roma, I-00185 Roma, Italy}
\affiliation{Mitchell Institute for Fundamental Physics and Astronomy, Texas A\&M University, College Station, Texas 77843, USA}
\affiliation{Istituto Nazionale di Fisica Nucleare Trieste, \ensuremath{^{qq}}Gruppo Collegato di Udine, \ensuremath{^{rr}}University of Udine, I-33100 Udine, Italy, \ensuremath{^{ss}}University of Trieste, I-34127 Trieste, Italy}
\affiliation{University of Tsukuba, Tsukuba, Ibaraki 305, Japan}
\affiliation{Tufts University, Medford, Massachusetts 02155, USA}
\affiliation{University of Virginia, Charlottesville, Virginia 22906, USA}
\affiliation{Waseda University, Tokyo 169, Japan}
\affiliation{Wayne State University, Detroit, Michigan 48201, USA}
\affiliation{University of Wisconsin, Madison, Wisconsin 53706, USA}
\affiliation{Yale University, New Haven, Connecticut 06520, USA}

\author{T.~Aaltonen}
\affiliation{Division of High Energy Physics, Department of Physics, University of Helsinki, FIN-00014, Helsinki, Finland; Helsinki Institute of Physics, FIN-00014, Helsinki, Finland}
\author{S.~Amerio\ensuremath{^{jj}}}
\affiliation{Istituto Nazionale di Fisica Nucleare, Sezione di Padova, \ensuremath{^{jj}}University of Padova, I-35131 Padova, Italy}
\author{D.~Amidei}
\affiliation{University of Michigan, Ann Arbor, Michigan 48109, USA}
\author{A.~Anastassov\ensuremath{^{v}}}
\affiliation{Fermi National Accelerator Laboratory, Batavia, Illinois 60510, USA}
\author{A.~Annovi}
\affiliation{Laboratori Nazionali di Frascati, Istituto Nazionale di Fisica Nucleare, I-00044 Frascati, Italy}
\author{J.~Antos}
\affiliation{Comenius University, 842 48 Bratislava, Slovakia; Institute of Experimental Physics, 040 01 Kosice, Slovakia}
\author{G.~Apollinari}
\affiliation{Fermi National Accelerator Laboratory, Batavia, Illinois 60510, USA}
\author{J.A.~Appel}
\affiliation{Fermi National Accelerator Laboratory, Batavia, Illinois 60510, USA}
\author{T.~Arisawa}
\affiliation{Waseda University, Tokyo 169, Japan}
\author{A.~Artikov}
\affiliation{Joint Institute for Nuclear Research, RU-141980 Dubna, Russia}
% ...

\collaboration{CDF Collaboration}
\altaffiliation[With visitors from]{
\ensuremath{^{a}}University of British Columbia, Vancouver, BC V6T 1Z1, Canada,
\ensuremath{^{b}}Istituto Nazionale di Fisica Nucleare, Sezione di Cagliari, 09042 Monserrato (Cagliari), Italy,
\ensuremath{^{c}}University of California Irvine, Irvine, CA 92697, USA,
\ensuremath{^{d}}Institute of Physics, Academy of Sciences of the Czech Republic, 182~21, Czech Republic,
\ensuremath{^{e}}CERN, CH-1211 Geneva, Switzerland,
\ensuremath{^{f}}Cornell University, Ithaca, NY 14853, USA,
\ensuremath{^{g}}University of Cyprus, Nicosia CY-1678, Cyprus,
\ensuremath{^{h}}Office of Science, U.S. Department of Energy, Washington, DC 20585, USA,
\ensuremath{^{i}}University College Dublin, Dublin 4, Ireland,
\ensuremath{^{j}}ETH, 8092 Z\"{u}rich, Switzerland,
\ensuremath{^{k}}University of Fukui, Fukui City, Fukui Prefecture, Japan 910-0017,
\ensuremath{^{l}}Universidad Iberoamericana, Lomas de Santa Fe, M\'{e}xico, C.P. 01219, Distrito Federal,
\ensuremath{^{m}}University of Iowa, Iowa City, IA 52242, USA,
\ensuremath{^{n}}Kinki University, Higashi-Osaka City, Japan 577-8502,
\ensuremath{^{o}}Kansas State University, Manhattan, KS 66506, USA,
\ensuremath{^{p}}Brookhaven National Laboratory, Upton, NY 11973, USA,
\ensuremath{^{q}}Queen Mary, University of London, London, E1 4NS, United Kingdom,
\ensuremath{^{r}}University of Melbourne, Victoria 3010, Australia,
\ensuremath{^{s}}Muons, Inc., Batavia, IL 60510, USA,
\ensuremath{^{t}}Nagasaki Institute of Applied Science, Nagasaki 851-0193, Japan,
\ensuremath{^{u}}National Research Nuclear University, Moscow 115409, Russia,
\ensuremath{^{v}}Northwestern University, Evanston, IL 60208, USA,
\ensuremath{^{w}}University of Notre Dame, Notre Dame, IN 46556, USA,
\ensuremath{^{x}}Universidad de Oviedo, E-33007 Oviedo, Spain,
\ensuremath{^{y}}CNRS-IN2P3, Paris, F-75205 France,
\ensuremath{^{z}}Universidad Tecnica Federico Santa Maria, 110v Valparaiso, Chile,
\ensuremath{^{aa}}The University of Jordan, Amman 11942, Jordan,
\ensuremath{^{bb}}Universite catholique de Louvain, 1348 Louvain-La-Neuve, Belgium,
\ensuremath{^{cc}}University of Z\"{u}rich, 8006 Z\"{u}rich, Switzerland,
\ensuremath{^{dd}}Massachusetts General Hospital, Boston, MA 02114 USA,
\ensuremath{^{ee}}Harvard Medical School, Boston, MA 02114 USA,
\ensuremath{^{ff}}Hampton University, Hampton, VA 23668, USA,
\ensuremath{^{gg}}Los Alamos National Laboratory, Los Alamos, NM 87544, USA,
\ensuremath{^{hh}}Universit\`{a} degli Studi di Napoli Federico I, I-80138 Napoli, Italy
}
\noaffiliation


\begin{abstract}
   We present an analysis of top-antitop quark production and
decay into a tau lepton, tau neutrino, and bottom quark using
data from $9 \, {\rm fb}^{-1}$ of integrated luminosity at the
Collider Detector at Fermilab.
   Dilepton events, where one lepton is an energetic electron
or muon and the other a hadronically-decaying tau lepton,
originating from proton-antiproton collisions at $\sqrt{s}
= 1.96 \, \mbox{TeV}$ are used. A top-antitop quark production
cross section of $8.1 \pm 2.1 \, {\rm pb}$ is measured, assuming
standard-model top-quark decays. By separately identifying for
the first time the single-tau and the ditau components, we
measure the branching fraction of the top quark into tau
lepton, tau neutrino, and bottom quark to be $(9.6 \pm 2.8)
\%$. The branching fraction of top-quark decays into a charged
Higgs boson and a bottom quark, which would imply violation of
lepton universality, is limited to be less than $5.9 \%$ at
$95 \%$ confidence level (for $\mathcal{B}({\rm H}^{-} \rightarrow
\tau \overline{\nu}) = 1$).
\end{abstract}

\pacs{14.65.Ha, 14.80.Da, 14.80.Fd}
\maketitle

The large integrated luminosity provided by the 2001-2011 operations
of the Tevatron proton-antiproton collider enables the Fermilab
collider detector experiments to perform precision measurements of
top-quark properties \cite{topprop}. Within the standard model the
top quark decays into a $\rm W$ boson and bottom quark, which is
the dominant decay mode \cite{PDG}. However, new particles beyond
the standard model, like charged bosons, could open additional decay
modes. Higgs bosons with unit electric charge are predicted by
extensions of the standard model that contain an extended Higgs
sector, such as two Higgs doublet models \cite{2HDM}. For charged Higgs
bosons lighter than the top quark, one of the most interesting decay
modes is into a tau lepton and neutrino, with a branching fraction
close to $1$ in a large region of parameter space.

This paper presents an analysis of dilepton events predominantly
originated from top-antitop-quark, $\rm t \overline{t}$, production,
where one of the leptons is an electron or muon and the other a
hadronically-decaying tau. For hadronically-decaying tau leptons,
both online event selection, trigger, and identification are
demanding due to the relatively short lifetime of tau leptons and
the presence of one or more neutrinos in the final state. We measure
top production and decay properties: the $\rm t \overline{t}$
production cross section and top branching fraction into tau lepton,
neutrino, and bottom quark. Separating top-quark decays where the
electron or muon originates from a tau decay from those where it
originates directly from a $\rm W$ boson decay enables us to measure
the branching fraction without relying on a theoretical cross-%
section calculation.

The Collider Detector at Fermilab, CDF, experiment \cite{CDFdetector}
is located at the Fermilab Tevatron. The data used in the analysis
presented here were collected between 2001 and 2011 at a center-of-%
mass energy of $1.96 \, \mbox{TeV}$ and correspond to an integrated
luminosity of $9 \, {\rm fb}^{-1}$. Of particular importance for the
analysis is the charged-particle trajectory measurement (tracking)
system. It is comprised of a silicon-microstrip-detector system
\cite{SVX} close to the collision region and an open-cell drift
chamber \cite{COT}, both immersed in a $1.4 \, \mbox{Tesla}$
solenoidal field. They enable measuring the transverse momentum
\cite{coordinate} of charged particles with a resolution of about $150
\, \mbox{MeV}/c$ at $10 \, \mbox{GeV}/c$. Outside the tracking system
are electromagnetic and hadronic sampling-calorimeters \cite{calorimeter}
segmented in a projective-tower geometry of about 0.1 unit in
pseudorapidity and 15 degrees in azimuthal angle \cite{coordinate}.
Drift chambers and
scintillators are located outside the calorimeters to identify muons
\cite{muon}. Events for this analysis are selected by triggers
designed to collect samples enriched in decays of low-momentum tau
leptons for non-standard-model physics searches. The triggers require
an electron or muon candidate with at least $8 \, \mbox{GeV}/c$ and a
charged particle of at least $5 \, \mbox{GeV}/c$ of transverse momentum.
In the later part of the data taking, triggers also require the track
to point to a narrow energy deposition of $5 \, \mbox{GeV}$ or more in
five or fewer calorimeter towers, thus requiring a more tau-like
signature.

At the Tevatron, $\rm t \overline{t}$ production via quark-anti\-quark
annihilation is the dominant top-quark production process. Top dilepton
events with one hadronic tau decay have two sources: (i) events with one
top quark decaying into an electron or muon, neutrino, and bottom quark
and the other into a tau lepton, neutrino, and bottom quark,
``single tau'', and (ii) events with both top quarks decaying into a tau
lepton, neutrino, and bottom quark and one tau decaying leptonically
and the other hadronically, ``ditau''. The neutrinos escape
direct detection and result in an momentum imbalance in the event. The
bottom quarks are each identified as cluster of hadronic energy, jet,
with a displaced secondary vertex, from the long-lived bottom-quark
hadrons.

After event reconstruction \cite{CDFdetector}, we select events with one
central, isolated electron or muon candidate of $10 \, \mbox{GeV}/c$ or
higher that meets standard CDF identification requirements \cite{CDFid}
and one central, isolated, one-prong or three-prong tau candidate
\cite{tauid} with transverse momentum $p_{\rm T} \ge 15$ and $20 \,
\mbox{GeV}/c$, respectively. The reconstruction and identification of a
hadronic tau decay starts with a calorimeter tower of transverse
energy $E_{\rm T} \ge 6 \, \mbox{GeV}$ \cite{coordinate} and a charged
particle of $p_{\rm T} \ge 6 \, \mbox{GeV}/c$ pointing to it.
Neighboring towers with $1 \, \mbox{GeV}$ or more of transverse energy
are added. Since tau decays yield highly collimated jets, calorimeter
clusters with more than six towers are rejected. A signal cone with
aperture $2 \vartheta$ is defined around the track. Its size
$\vartheta = \mbox{min}(0.17, \: 5.0 \, \mbox{GeV} / E)$ depends
on the calorimeter-cluster energy $E$, i.e. shrinks for cluster
energies above $30 \, \mbox{GeV}$. Charged particles with $p_{\rm T} \ge
1 \, \mbox{GeV}/c$ within the cone are associated with the tau candidate
and define its track multiplicity or number of prongs. For one-prong
tau candidates, the calorimeter cluster is required to have $E_{\rm T}
\ge 10 \, \mbox{GeV}$ and for three-prong candidates $E_{\rm T} \ge 15 \,
\mbox{GeV}$. To suppress contamination from minimum ionizing particles
in the calorimeter,
the $E_{\rm T}$ of the calorimeter cluster should be at least $40 \%$
of the scalar sum of the transverse momenta of charged particles in the
signal cone. An isolation annulus between the signal cone and $\pi / 6$
is used to further suppress quark and gluon jets. No charged particle
with $p_{\rm T} \ge 1.5 \, \mbox{GeV}/c$ or neutral pion candidate
\cite{pi0} is allowed in this region and the scalar transverse momentum
sum of the charged particles inside the region must not exceed $2 \,
\mbox{GeV}/c$. To suppress electron contamination, tau candidates
associated with very small energy in the hadronic calorimeter
compared to the transverse momentum sum of the charged particles are
rejected \cite{xiprime}. The four-momentum vector of the tau candidate
is calculated from the charged particles and neutral pions inside the
signal cone. Since the tau decay produces a neutrino, the reconstructed
mass should be smaller than the mass of the tau lepton and it is thus
required to be less than $1.8 \, \mbox{GeV}/c^2$. Tau candidates whose three
final-state charged particles have the same electric charge are rejected
as are events where the charge of the tau candidate and of the electron
or muon candidate have the same sign.

Identified ${\rm Z}^{0}$ decays and low-mass muon pairs from the
Drell-Yan process are removed with dilepton mass requirements. If the
two-body mass of the electron plus any calorimeter cluster with over
$90 \%$ of energy in the electromagnetic compartment falls within the
range of $86$ to $96 \, \mbox{GeV}/c^2$, the event is rejected.
Similarly, if the two-body mass of the muon and any minimum ionizing
particle falls within the range of $76$ to $106 \, \mbox{GeV}/c^2$ or
is below $15 \, \mbox{GeV}/c^2$ the event is rejected.

For the identification of bottom-quark jets, we use the {\sc SECVTX}
tagger \cite{CDFsim} to identify secondary vertices within a jet that
are displaced from the primary interaction vertex of the event. We
require two jets \cite{jetclu}, one with $E_{\rm T} > 20 \, \mbox{GeV}$
and a second with $E_{T\rm } > 15 \, \mbox{GeV}$, both within $| \eta |
\le 2$ \cite{coordinate}. At least one jet should be tagged as a
candidate for containing a long-lived bottom-quark hadron, i.e., have
a secondary vertex with a decay length in the transverse plane that
exceeds three times its uncertainty. The efficiency of the {\sc SECVTX}
tagger is about $44 \%$ for bottom-quark jets in the central region
with about a $1 \%$ mistag rate for light-quark and gluon jets.

Events from $\rm t \overline{t}$ production with leptonic $\rm W$
decays and one hadronic tau decay have three or five neutrinos in the
final state (not counting possible semileptonic heavy-quark decays).
The sum of the transverse momenta of the neutrinos is
measured by the missing transverse energy ${\not\!\!\!\,E_{\rm T}}$
\cite{missingET} of the event. Events are required to have
${\not\!\!\!\,E_{\rm T}} \ge 10 \, \mbox{GeV}$, a small amount
compared to other $\rm t \overline{t}$ analyses \cite{topprop}
with one (or two) neutrinos from a $\rm W$ decay in the final state.

To purify the selection of $\rm t \overline{t}$ events we exploit the
large top-quark mass and require events to have a large amount of
transverse energy in the final state: $H_{\rm T} = {\not\!\!\!\,E_{\rm T}} +
E_{\rm T}^{\mbox{tau}} + \sum E_{\rm T}^{\mbox{jet}} \ge 150 \,
\mbox{GeV}$ ($\ge 155 \, \mbox{GeV}$ in case of three-prong tau
candidates). Contrary to other top quark analyses \cite{topprop}, the
$E_{\rm T}$ of the electron or muon is not included in $H_{\rm T}$.
The electron or muon can come from either top or tau decay and
including its $E_{\rm T}$ in the $H_{\rm T}$ would favor single-tau
(with a more energetic electron or muon) over ditau events in the
selection. This selection defines the initial analysis sample.

To study background sources and the signal we generate events using
the {\sc Alpgen} and {\sc Pythia} Monte Carlo programs \cite{MCgen},
using a top quark mass of $173 \, \mbox{GeV}/c^2$. To mimic geometrical
and kinematical acceptances, the generated events are passed through
a {\sc Geant}-based detector simulation \cite{CDFsim}. Event yields
are normalized using theoretical next-to-leading-order cross-section
calculations \cite{dibosonNLO} and scaled for trigger efficiencies
and any differences in lepton identification efficiencies between
data and simulated events.

To estimate the background contribution from quark and gluon jets
misidentified as hadronic tau decays, events triggered on single jets
or on single electrons or muons are used. First, the probability of a
hadronic jet to be misidentified as a tau lepton is measured. In the
second step, this probability is used to weight the jets in events with an
electron or muon candidate to estimate and study the misidentified-%
tau-lepton contribution in our electron-or-muon-plus-tau sample. Signal
events are naturally excluded in this method due to the exclusive one-%
tau-candidate selection in the initial analysis sample.

For the misidentification-probability measurement, we use jet-triggered
data with calorimeter jet-$E_{\rm T}$ thresholds of $5$ to $100 \,
\mbox{GeV}$.  The trigger bias is removed by requiring at least two jets
in each event to pass the trigger requirements. For the probability
calculation we use tau-like jets, i.e., calorimeter jets that
pass the tau identification requirements above except for the isolation
and mass requirements. There is a small contribution from genuine tau-%
lepton decays in the jet sample due to vector boson (and top-quark)
decays involving a tau. This contribution is suppressed by requiring
the jet events to have insignificant missing transverse energy
\cite{topditaupublicnote} and no identified electron or
muon with $E_{\rm T} > 10 \, \mbox{GeV}$. Leading and subleading jets
have different proportions of gluons and quarks resulting in
different probabilities for tau misidentification. We treat these
jets separately, average their misidentification rates to determine the
nominal tau-misidentification probability, and use the difference to
each as a measure of the systematic uncertainty. The probability is
parametrized as a function of jet $E_{\rm T}$, $\eta$, and number of
prongs. Probabilities range from $1 \%$ to $10 \%$ per tau-like jet
or $0.1 \%$ to $1 \%$ per generic jet. The measurement of the tau-%
misidentification probability is checked in two control regions,
a multi-jet and a $\rm W$ plus jet enriched sample.

To estimate the contribution of jets misidentified as tau candidates
in this analysis, single-lepton-triggered data are used. CDF recorded
events requiring only a $p_{\rm T} > 8 \, \mbox{GeV}/c$ electron or
muon candidate. For each tau-like jet in those events, we mimic a tau
candidate with the above measured probability, correct for the
difference in sample size with respect to the signal sample, and
analyze it analogously to the signal sample. With this approach we
simulate event characteristics, study the contribution of events with
a misidentified tau lepton, and properly apply kinematic selections
to reduce this background.

\begin{figure}[htb]
\centering
\includegraphics[width=0.48\textwidth]{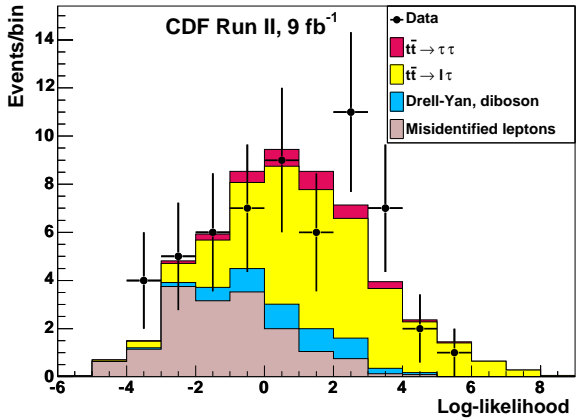}
\caption{Log-likelihood $\mathcal{L}_{1}$ distribution for data, top
dilepton, Drell-Yan plus diboson production, and misidentified lepton
events.}
\label{likelihood1}
\end{figure}

In the initial analysis sample we observe 58 events with an
expectation of about 34 $\rm t \overline{t}$ dilepton events
\cite{ttbar_crosssection} and about 4 events from Drell-Yan and
diboson production. Astoundingly, the calculation of misidentified
tau leptons shows a contribution of approximately 16 events. We find
over half of them to be $\rm t \overline{t}$ events, but with one
of the top quarks decaying leptonically and the other hadronically.
Four jets provide the occasion to be misidentified as a tau lepton,
while the rest of the event satisfies the selection with effectively
the same efficiency as the signal. To reject this background, a
likelihood function $\mathcal{L}_{1}$ is constructed based on two
tau identification variables and three kinematic variables to
distinguish single-lepton from dilepton-$\rm t \overline{t}$ events:
(i) The ratio of tau-calorimeter cluster $E_{\rm T}$ to signal-cone-%
charged-particle-$p_{\rm T}$ sum is used in the tau identification
to reject muon background. Compared to genuine tau leptons the
distribution of this ratio for jets misidentified as taus is broader
with a long tail. (ii) The charged-particle $p_{\rm T}$ sum in the
isolation annulus is required to be less than $2 \, \mbox{GeV}/c$
in the tau candidate selection. It provides further discrimination
as tau leptons in top quark decays are isolated, while jets
misidentified as taus have a nearly uniform distribution within the
remaining phase space. (iii) The one neutrino in the single-%
lepton $\rm t \overline{t}$ events yields a ${\not\!\!\!\,E_{\rm T}}$
of around half the $\rm W$ boson mass and (iv) a transverse mass of
electron or muon plus missing $E_{\rm T}$ up to the $\rm W$ mass. (v)
Events with a hadronic top-quark decay have an extra jet. One of the
two extra jets in the event must be misidentified as a tau lepton.
The jet may not be reconstructed, and dilepton $\rm t \overline{t}$
events may have additional jets from initial- or final-state
radiation. The $E_{\rm T}$ of any third jet becomes the final variable
for the likelihood. Figure~\ref{likelihood1} shows the distribution of
the likelihood function for data compared to the expectation from top
dilepton, Drell-Yan, diboson, and misidentified tau lepton events. A
$\log \mathcal{L}_{1} > 0$ requirement leaves 36 events in the data
with $26.7 \pm 3.4$ $\rm t \overline{t}$ dilepton events, $3.1 \pm 0.5$
Drell-Yan and diboson events, and $4.0 \pm 1.1$ events expected from
jets misidentified as tau leptons.

Assuming a standard-model top-quark decay, i.e., a branching fraction
$\mathcal{B}({\rm t \rightarrow W^{+} b}) = 1$, an acceptance corrected
$\rm t \overline{t}$ production cross section of $8.1 \pm 1.7 \mbox{(stat)}
^{+1.2}_{-1.1} \mbox{(syst)} \pm 0.5 \mbox{(lumi)} \, {\rm pb}$ is
measured. Acceptance corrections are based on leading order parton
shower Monte Carlo simulations and the CDF detector simulation described
above. Branching fractions for both $\rm W$ boson and tau lepton decays are
taken from \cite{PDG}. Lepton, jet, and bottom-quark-tagging efficiencies
are estimated previously \cite{topprop,CDFid,tauid,CDFsim}. The systematic
uncertainties include experimental contributions from lepton acceptance
and identification, trigger efficiency, tau- and jet-energy-scale
corrections, tagging efficiency ($\pm 5 \%$), mistag rate ($\pm 20 \%$),
tau-misidentification probability ($^{+\ 7 \%}_{-20 \%}$)
\cite{topditaupublicnote}, modeling of additional collisions contained
in an event, pile-up, and measurement of the integrated luminosity
($\pm 5.9 \%$) and theoretical contributions from the cross section of
Drell-Yan and diboson production, choice of parton distribution
functions, color reconnection, initial- and final-state radiation
($\pm 9 \%$), fragmentation and parton showering. The uncertainty of
the dominant systematic effects are provided in parenthesis (which is
not their contribution to the cross section uncertainty). This
cross section measurment in the ditau channel complements the more
precise measurements in the electron, muon and hadronic channels
\cite{topprop}.

\begin{figure}[htbp]
\centering
\includegraphics[width=0.48\textwidth]{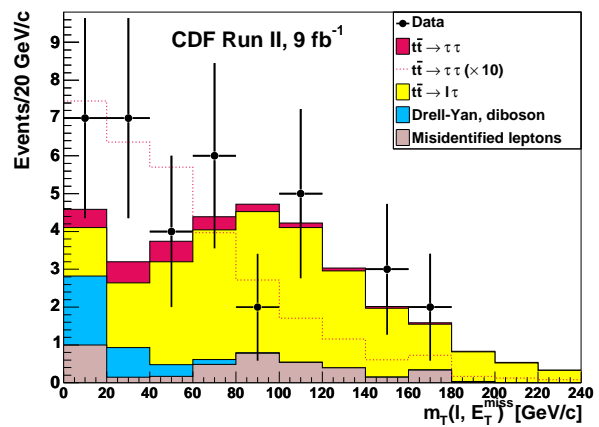}
\caption{Transverse mass distribution of the electron or muon plus
missing $E_{\rm T}$ for data, single-tau $\rm t \overline{t}$,
ditau $\rm t \overline{t}$, and non-$\rm t \overline{t}$ events.
For the purpose of comparing the single-tau and ditau shapes, used
in the likelihood, a line shows the ditau contribution scaled to
the number of single-tau events and plotted on top of the background.}
\label{l2_mass_t}
\end{figure}

\begin{figure}[htbp]
\centering
\includegraphics[width=0.48\textwidth]{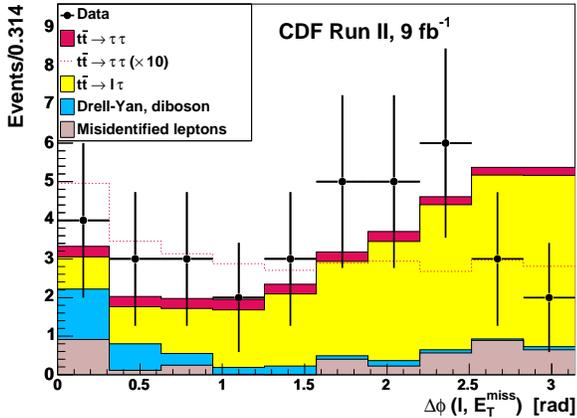}
\caption{Distribution of the azimuthal angle between electron or muon
and missing $E_{\rm T}$ for data, single-tau $\rm t \overline{t}$,
ditau $\rm t \overline{t}$, and non-$\rm t \overline{t}$ events. For
the purpose of comparing the single-tau and ditau shapes, used in the
likelihood, a line shows the ditau contribution scaled to the number
of single-tau events and plotted on top of the background.}
\label{l2_delta_phi}
\end{figure}

\begin{figure}[htbp]
\centering
\includegraphics[width=0.48\textwidth]{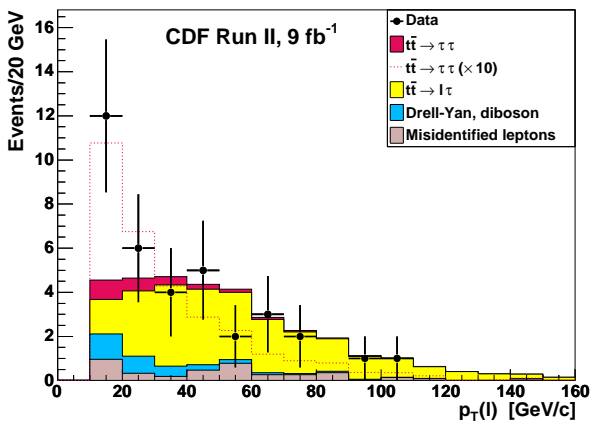}
\caption{Transverse momentum distribution of the electron or muon for
data, single-tau $\rm t \overline{t}$, ditau $\rm t \overline{t}$, and
non-$\rm t \overline{t}$ events. For the purpose of comparing the
single-tau and ditau shapes, used in the likelihood, a line shows the
ditau contribution scaled to the number of single-tau events and
plotted on top of the background.}
\label{l2_lep_p_t}
\end{figure}

\begin{figure}[htbp]
\centering
\includegraphics[width=0.48\textwidth]{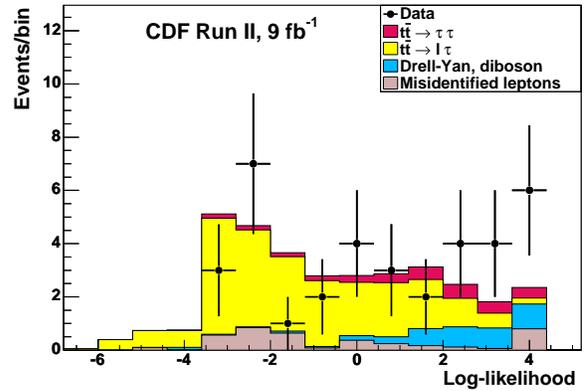}
\caption{Log-likelihood $\mathcal{L}_{2}$ distribution for data, top
dilepton, Drell-Yan plus diboson production, and misidentified lepton
events as determined by the fit.}
\label{likelihood2}
\end{figure}

Using the theoretical $\rm t \overline{t}$ production cross section
instead of the standard-model decay branching fractions, we can extract
a branching fraction of the top-quark decay into tau lepton, tau
neutrino, and bottom quark. However, the data sample contains two
$\rm t \overline{t}$ components: a single-tau component that is
proportional to the top quark into tau lepton, tau neutrino, and
bottom-quark branching fraction (times the top quark into electron
or muon, neutrino, and bottom-quark branching fraction) and a ditau
component that is proportional to the branching fraction squared.
Separating the two components allows measurement of the top quark
into tau, neutrino, and bottom-quark branching fraction (and the $\rm
t \overline{t}$ production cross section) directly without theoretical
assumption on either. For this, a second likelihood,
$\mathcal{L}_{2}$, is constructed using the following three variables.
The leptonic tau decay yields two neutrinos close to the electron
or muon direction, in addition to the neutrino from the $\rm W$ decay.
(i) This impacts the distribution of transverse mass \cite{coordinate}
of the electron or muon plus missing $E_{\rm T}$ (Fig.~\ref{l2_mass_t})
and (ii) the distribution of the azimuthal angle between electron or
muon and missing $E_{\rm T}$ (Fig.~\ref{l2_delta_phi}). (iii) It also
leads to, on average, lower $p_{\rm T}$ of electrons or muons in ditau
events, as shown in Fig.~\ref{l2_lep_p_t}. The ditau component
contributes more at large $\mathcal{L}_{2}$ while the single-tau
component is shifted toward smaller values of $\mathcal{L}_{2}$, as
shown in Fig.~\ref{likelihood2}.

   We use the MClimit \cite{MClimit} package to fit the likelihood
distribution with a single-tau component that has a linear dependence
on the branching fraction of top quark into tau, neutrino, and
bottom-quark, and a ditau component that has a quadratic branching-%
fraction dependence. The expected background contributions from
Drell-Yan, diboson production, and jets misidentified as tau leptons
are included in the fit and allowed to vary within their uncertainties.
The systematic uncertainties on the event yield are included via
nuisance parameters. We check the effect of the largest systematic
uncertainty (from the probability distribution of jets being
misidentified as tau leptons) on the shape of the likelihood
distribution and found it to be small. For the most precise result
we make use of a third branching fraction dependency by including
the $\rm t \overline{t}$ production cross-section measured by CDF in
the single-lepton channel with its uncertainty \cite{topsinglelep}
as a constraint in the $\chi^{2}$ fit \cite{BRdependency} and obtain
   \[ \mathcal{B}({\rm t \rightarrow \tau \nu b}) = 0.096 \pm 0.028, \]
with a single-tau component of $22.7$ events and ditau component of
$3.1$ events (Fig.~\ref{likelihood2}). The ratio of leptonic top
branching ratios can be derived from this, and we find $\mathcal{B}({\rm
t \rightarrow \tau \nu b}) / ( \, ( \mathcal{B}({\rm t \rightarrow e \nu b})
+ \mathcal{B}({\rm t \rightarrow \mu \nu b}) ) /2 \, ) = 0.88 \pm 0.26$,
confirming lepton universality in top quark decays.

At $95 \%$ confidence level the fit excludes a branching fraction
of $15.8 \% $ or more of top quarks decaying into tau lepton, tau
neutrino, and bottom quark (using the likelihood-ratio method
\cite{PDG}). A previous CDF analysis \cite{Eusebi} found the
acceptance for $\rm t \overline{t}$ tau events with decay via
charged Higgs boson to be equal to the acceptance of events with
decay via $\rm W$ boson for charged-Higgs masses between $80$ and
$140 \, \mbox{GeV}/c^{2}$. Assuming that the top quark decays only
into $\rm W$ and charged Higgs bosons and $\mathcal{B}({\rm H}^{-}
\rightarrow \tau \overline{\nu}) = 1$, we exclude branching fractions
for top quark decays into a charged Higgs boson and a bottom quark of
$5.9 \%$ or more at $95 \%$ confidence level using the likelihood-ratio
ordering \cite{Feldman_Cousins}.

With the discovery of a neutral Higgs boson at $125 \, \mbox{GeV}/c^{2}$
by the ATLAS and CMS experiments \cite{Higgs}, the question of an
extended Higgs sector gains interest. Previous searches for a charged
Higgs boson at LEP constrained its mass to be $m_{\rm H^{\pm}} > 78.6 \,
\mbox{GeV}/c^{2}$ \cite{LEPcHiggs} and at the Tevatron constrained
the branching fractions to be
$\mathcal{B}({\rm t \rightarrow H^{+} b}) < 0.15 - 0.19$ (for
$\mathcal{B}({\rm H}^{-} \rightarrow \tau \overline{\nu}) = 1$) and
$\mathcal{B}({\rm t \rightarrow H^{+} b}) < 0.10 - 0.30$
(for $\mathcal{B}({\rm H^{+} \rightarrow c \overline s}) = 1$)
\cite{TEVcHiggs}. Recent LHC searches improved the limits to
$\mathcal{B}({\rm t \rightarrow H^{+} b}) < 0.01 - 0.05$ \cite{AtlascHiggs}
and $< 0.02 - 0.03$ \cite{CMScHiggs} for $\mathcal{B}({\rm H}^{-}
\rightarrow \tau \overline{\nu}) = 1$.

In summary, we present an analysis of dilepton $\rm t \overline{t}$
events using $9 \, {\rm fb}^{-1}$ of integrated luminosity at CDF.
The analysis separates the single and ditau components for the first
time, measuring the branching fraction of the top-quark decay into
tau lepton, tau neutrino, and bottom quark at $(9.6 \pm 2.8) \%$, and
testing lepton universality of the decay. The limit on the branching
fraction of top quark into charged Higgs boson and bottom quark is
comparable with recent measurements in proton-proton collisions
\cite{AtlascHiggs, CMScHiggs}.

We thank the Fermilab staff and the technical staffs of the participating
institutions for their vital contributions. This work was supported by the
U.S.\ Department of Energy and National Science Foundation; the Italian
Istituto Nazionale di Fisica Nucleare; the Ministry of Education, Culture,
Sports, Science and Technology of Japan; the Natural Sciences and
Engineering Research Council of Canada; the National Science Council of
the Republic of China; the Swiss National Science Foundation; the
A.P.\ Sloan Foundation; the Bundesministerium f\"ur Bildung und Forschung,
Germany; the Korean World Class University Program, the National Research
Foundation of Korea; the Science and Technology Facilities Council and
the Royal Society, UK; the Russian Foundation for Basic Research; the
Ministerio de Ciencia e Innovaci\'{o}n, and Programa Consolider-Ingenio
2010, Spain; the Slovak R\&D Agency; the Academy of Finland; the Australian
Research Council (ARC); and the EU community Marie Curie Fellowship
contract 302103.


\begin{thebibliography}{99}
\bibitem{coordinate}
   CDF uses a right-handed coordinate system with origin at the nominal
   interaction point. $\theta$ and $\phi$ are the polar and azimuthal
   angles, defined with respect to the $z$-axis, the proton beam direction.
   The transverse momentum of a particle is $p_{\rm T} = p \sin ( \theta
   )$ and the pseudorapidity $\eta = - \ln \tan ( \theta / 2 )$.
   Similarly for the transverse energy, $E_{\rm T}$. The transverse
   mass of two particles is calculated as $m_{\rm T} = \sqrt{ 2
   E_{\rm T}^{\rm 1} E_{\rm T}^{\rm 2} (1 - \cos(\phi^{\rm 1} -
   \phi^{\rm 2}))}$.
\bibitem{pi0}
   Neutral pion candidates are energy deposits in the electromagnetic
   calorimeter and are not reconstructed two photons at the $\pi ^{0}$
   mass.
\bibitem{xiprime}
   For best electron and muon rejection, the fraction of energy in the
   electromagnetic compartment of the calorimeter is actually compared
   to the ratio of calorimeter energy and momentum sum of the charged
   particles in signal cone, $( E / \sum | \vec{p} | ) \: (0.95 -
   E^{\mbox{em}} / E) > 0.1$.
\bibitem{missingET}
   While proton and antiproton have equal but opposite momenta, the
   interacting partons likely have different momenta. Therefore the
   interaction has an unknown longitudinal boost.
   The missing transverse energy is defined as the opposite of the
   transverse component of the vector sum of the energy in the
   calorimeters towers. We use ${\not\!\!\!\,E_{\rm T}}$ to denote
   the scalar magnitude of the missing $E_{\rm T}$.
   The energy of identified minimum-ionizing particles (muons) is
   included in the missing $E_{\rm T}$ calculation and the energy of
   jets is corrected for the calorimeter response and absolute scaling.
\bibitem{ttbar_crosssection}
   For the expected number of $\rm t \overline{t}$ events, a top-quark
   mass of ${\rm m_{t}} = 173 \, \mbox{GeV}/c^{2}$ and cross section
   of $\sigma_{\rm t \overline{t}} = 7.5 \, {\rm pb}$ is used.
\bibitem{BRdependency}
   The cross-section measurement in the single-lepton channel has a
   branching-fraction dependence of $\mathcal{B}^{\text{SM}}({\rm t
   \rightarrow e \nu b}) \times [ 1 + ( \mathcal{B}^{\text{SM}}({\rm t
   \rightarrow \tau \nu b}) - \mathcal{B}^{\text{fit}}({\rm t
   \rightarrow \tau \nu b}) ) / ( 1 - \mathcal{B}^{\text{SM}}
   ({\rm t \rightarrow \tau \nu b}) ) ]$, which is used in the fit.
\end{thebibliography}
\end{document}